\documentclass[10pt]{article}
\usepackage{a4}
\usepackage{epsfig}
\usepackage[T2A]{fontenc}
\usepackage[english]{babel}

\begin{document}
\newcommand\Tr{{~\rm Tr~}}

\title{The spontaneous generation of magnetic fields at
high temperature in $SU(2)$-gluodynamics on a lattice}
\author{V.Demchik\footnote{vadimdi@yahoo.com}, V.Skalozub\footnote{skalozub@ff.dsu.dp.ua}
\\
{\it{Dniepropetrovsk National University, 49025 Dniepropetrovsk,
Ukraine}}}
\date{\empty}
\maketitle
\begin{abstract}
The spontaneous generation of the chromomagnetic field at high
temperature is investigated in a lattice formulation of the
$SU(2)$-gluodynamics. The procedure of studying this phenomenon is
developed. The Monte Carlo simulations of the free energy on the
lattices $2 \times 8^3$, $2\times 16^3$ and $4 \times 8^3$ at
various temperatures are carried out. The creation of the field is
indicated by means of the $\chi^2$-analysis of the data set
accumulating 5-10 millions MC configurations. A comparison with
the results of other approaches is done.
\end{abstract}

\section{Introduction}
Among interesting problems of modern cosmology the origin of
large-scale magnetic fields is intensively attacked nowadays.
Various mechanisms of the field generation at different stages of
the universe evolution were proposed \cite{GR}. Basically they are
grounded on the idea of Fermi, Chandrasekhar and Zel'dovich that
to have the present day galaxy magnetic fields of order $\sim 1\mu
G$ correlated on a scale $\sim 1$Mpc seed magnetic fields must be
present in the early universe. These fields had been frozen in a
cosmic plasma and then amplified by some of the mechanisms of the
field amplification. One of the ways to produce seed fields is a
spontaneous vacuum magnetization at high temperature $T$
\cite{EO,SZ,SB,PL}. Actually, this is an extension of the Savvidy
model for the QCD vacuum \cite{SG}, proposed already at $T = 0$
and describing the creation of the Abelian chromomagnetic fields
due to a vacuum polarization, in case of nonzero temperature. At
zero temperature this field configuration is unstable because of
the tachyonic mode in the gluon spectrum. At $T\not = 0$, the
possibility of having strong temperature-dependent and stable
magnetic fields was discovered \cite{SB}. The field stabilization
is ensured by the temperature and field dependent gluon magnetic
mass.

Another related field of interest is the deconfinement phase of
QCD. As it was realized recently, this is not the gas of free
quarks and gluons, but a complicate interacting system of them.
This was discovered at RHIC experiments \cite{RHIC1} and observed
in either perturbative \cite{SB,S2} or nonperturbative \cite{AG}
investigations of the vacuum state with magnetic fields at high
temperature. In Refs. \cite{SB,S2} the spontaneous creation of the
chromomagnetic fields of order $gB \sim g^4 T^2$ was observed in
$SU(2)$- and $SU(3)$-gluodynamics within the one-loop plus daisy
resummation accounted for. In Ref. \cite{AG} the chromomagnetic
condensate of same order was obtained in stochastic QCD vacuum
model and method of dimensional reduction by comparison with
lattice data. In Refs. \cite{CC} the response of the vacuum to the
influence of strong external fields at different temperatures has
been investigated and it was shown that the confinement is
restored by increasing the strength of the applied field. These
results stimulated the present investigation.

We are going to determine the spontaneous creation of magnetic
fields in lattice simulations of $SU(2)$-gluodynamics. In contrast
to the problems in the external field, in the case of interest the
field strength is a dynamical variable which values at different
temperatures have to be determined by means of the minimization of
the free energy. This procedure is not a simple one as in
continuum because the field strength on a lattice is quantized. To
deal with this peculiarity, we consider magnetic fluxes on a
lattice as the main objects to be investigated. The fluxes take
continuous values, and therefore the minimization of the free
energy in presence of magnetic field can be fulfilled in a usual
way. These speculations serve as an explanation of the strategy of
our calculations.

One of the methods to introduce a magnetic flux on a lattice is to
use the twisted boundary conditions (t.b.c.) \cite{TH}. In this
approach the flux is a continuous quantity. So, in what follows we
consider the free energy $F(\varphi)$ with the magnetic flux
$\varphi$ on a lattice in the $SU(2)$-gluodynamics and calculate
its values at different temperatures by means of Monte Carlo (MC)
simulations. We will show that the global minimum of $F(\varphi)$
is located at some non-zero value $\varphi_{min}$ dependent on the
temperature. It means the spontaneous creation of the
temperature-dependent magnetic fields in the deconfinement phase.

The paper is organized as follows. In sect. 2 some necessary
information about the magnetic fluxes on a lattice is adduced. In
sect. 3 the calculation details and the results are given. Section
4 is devoted to discussion.

\section{Magnetic fields on a lattice}
In perturbation theory, the value of the macroscopic (classical)
magnetic field generated inside a system is determined by the
minimization of the free energy functional. The interaction with
the classical field is introduced by splitting the gauge field
potential in two parts: $A_\mu=\bar{A_\mu}+A_\mu^R$, where
$A_\mu^R$ describes a radiation field and
$\bar{A_\mu}=(0,0,Hx^1,0)$ corresponds to the constant magnetic
field directed along the third axis. However, on a lattice, the
direct detection of the spontaneously generated field strength by
straightforward analysis of the configurations, which are produced
in the MC simulations, seems to be problematic. Therefore, it is
reasonable to follow the approach used in the continuum field
theory.

First, let us write down the free energy density,
\begin{eqnarray}
\label{freeen}
F(\varphi)=-\log\frac{Z(\varphi)}{Z(0)}, \\
Z(\varphi)=\int [DU(\varphi)] \exp\{-S(U(\varphi))\}.
\end{eqnarray}
Here, $Z(\varphi)$ and $Z(0)$ are the partition function at finite
and zero magnetic fluxes, respectively; the link variable $U$ is
the lattice analogue of the potential $A_\mu$.

The free energy density relates to the effective action as
follows,
\begin{eqnarray}
\label{enactions} F(\varphi)=\bar{S}(\varphi)-\bar{S}(0),
\end{eqnarray}
where $\bar{S}(\varphi)$ and $\bar{S}(0)$ are the effective
lattice actions with and without magnetic field, correspondingly.

To detect the spontaneous creation of the field it is necessary to
show that the free energy density has the global minimum at a
non-zero magnetic flux, $\varphi_{min}\not =0$.

In what follows, we use the hypercubic lattice $L_t\times L_s^3$
($L_t<L_s$) with the hypertorus geometry; $L_t$ and $L_s$ are the
temporal and the spatial sizes of the lattice, respectively. In
the limit of $L_s \to \infty$ the temporal size $L_t$ is related
to physical temperature. The one-plaquette action of the $SU(2)$
lattice gauge theory can be written as
\begin{eqnarray}
\label{Wilson} S_W=\beta\sum_x\sum_{\mu>\nu}\left[1-\frac12 \Tr
U_{\mu\nu}(x)\right];\\\label{plaq}
U_{\mu\nu}(x)=U_\mu(x)U_\nu(x+a\hat{\mu})U_\mu^\dagger(x+a\hat{\nu})U_\nu^\dagger(x),
\end{eqnarray}
where $\beta=4/g^2$ is the lattice coupling constant, $g$ is the
bare coupling, $U_\mu(x)$ is the link variable located on the link
leaving the lattice site $x$ in the $\mu$ direction,
$U_{\mu\nu}(x)$ is the ordered product of the link variables.

The effective action $\bar{S}$ in (\ref{enactions}) is the Wilson
action averaged over the Boltzmann configurations, produced in the
MC simulations.

The lattice variable $U_\mu(x)$ can be decomposed in terms of the
unity, $I$, and Pauli, $\sigma_j,$ matrices in the color space,
\begin{eqnarray}
U_\mu(x)=IU_\mu^0(x)+i\sigma_jU_\mu^j(x)=\left(\begin{array}{cc}
U_\mu^0(x)+iU_\mu^3(x) &
U_\mu^2(x)+iU_\mu^1(x)\\
-U_\mu^2(x)+iU_\mu^1(x) &
U_\mu^0(x)-iU_\mu^3(x)\end{array}\right).
\end{eqnarray}
The four components $U_\mu^j(x)$ are subjected to the
normalization condition $\sum_j U_\mu^j(x)U_\mu^j(x)=1$. Hence,
only three components are independent.

Since the spontaneously generated magnetic field is to be the
Abelian one, the Abelian parametrization of the lattice variables
is used to introduce the magnetic field,
\begin{eqnarray}
\label{param2}
U_\mu(x)=\left(\begin{array}{cc}\cos\phi_\mu(x)e^{i\theta_\mu(x)}
& \sin \phi_\mu(x)e^{i\chi_\mu(x)}\\
-\sin\phi_\mu(x)e^{-i\chi_\mu(x)} &
\cos\phi_\mu(x)e^{-i\theta_\mu(x)}\end{array}\right),
\end{eqnarray}
where the angular variables are changed in the following ranges
$\theta,\chi\in[-\pi;+\pi)$, $\phi\in[0;\pi/2)$.

The Abelian part of the lattice variables is represented by the
diagonal components of the matrix and the condensate Abelian
magnetic field influences the field $\theta_\mu(x)$, only.

The second important task is to incorporate the magnetic flux in
this formalism. The most natural way was proposed by 't Hooft
\cite{TH}. In his approach, the constant homogeneous external flux
$\varphi$ in the third spatial direction can be introduced by
applying the following t.b.c.:
\begin{eqnarray}
& &U_\mu(L_t,x_1,x_2,x_3)=U_\mu(0,x_1,x_2,x_3),\\\nonumber &
&U_\mu(x_0,L_s,x_2,x_3)=U_\mu(x_0,0,x_2,x_3),\\\nonumber &
&U_\mu(x_0,x_1,L_s,x_3)=e^{i\varphi}
U_\mu(x_0,x_1,0,x_3),\\\nonumber &
&U_\mu(x_0,x_1,x_2,L_s)=U_\mu(x_0,x_1,x_2,0).
\end{eqnarray}
It could be seen, the edge links in all directions are identified
as usual periodic boundary conditions except for the links in the
second spatial direction, for which the additional phase $\varphi$
is added (Fig. 1). In the continuum limit, such t.b.c. settle the
magnetic field with the potential $A_\mu(x)=(0,0,Hx^1,0)$. The
magnetic flux $\varphi$ is measured in angular units and can take
a value from $0$ to $2\pi$.
\begin{figure}
\begin{center}
\includegraphics[bb=58 526 324 779,width=0.35\textwidth]{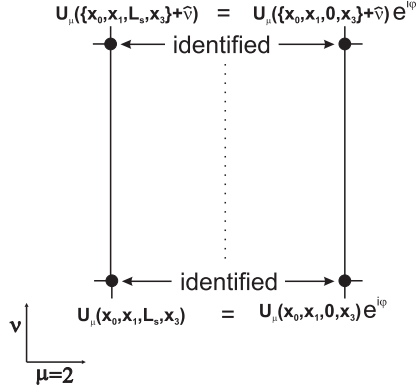}
\end{center}
\caption{The plaquette presentation of the twisted boundary
conditions.}
\end{figure}

The lattice variables (in the Abelian parametrization) in the
presence of the magnetic flux $\varphi$ are
\begin{eqnarray}
\label{paramphi}
U_\mu(x)=\left(\begin{array}{cc}\cos\phi_\mu(x)e^{i(\theta_\mu(x)+\varphi_\mu(x))}
& \sin \phi_\mu(x)e^{i\chi_\mu(x)}\\
-\sin\phi_\mu(x)e^{-i\chi_\mu(x)} &
\cos\phi_\mu(x)e^{-i(\theta_\mu(x)+\varphi_\mu(x))}\end{array}\right),
\end{eqnarray}
where $\varphi_\mu(x)=\varphi$ for the edge links at
$x=(x_0,x_1,L_s,x_3)$ with $\mu=2$ and $\varphi_\mu(x)=0$ for
other links.

The total flux through the plane spanned by the plaquettes $p$,
which affects the edge links at $x=(x_0,x_1,L_s,x_3)$ with
$\mu=2$, is
\begin{eqnarray}
\label{totflux}
&&g\Phi=\sum_{p\in plane} (\theta_p+\varphi), \\
&&\theta_p=\theta_{\mu\nu}(x)=\theta_\mu(x)+\theta_\nu(x+a\hat{\mu})-\theta_\mu(x+a\hat{\nu})-\theta_\nu(x).
\end{eqnarray}
Eq. (\ref{totflux}) is the lattice analogue of the flux in the
continuum:
\begin{eqnarray}
\Phi_c=\int_S d^2\sigma_{\mu\nu}F_{\mu\nu}.
\end{eqnarray}
In this approach the variable $\varphi$ describes a flux through
the whole lattice plane, not just through an elementary plaquette.

The t.b.c. for the components (\ref{paramphi}),
\begin{eqnarray}
U_\mu^0(x)&=&\cos(\theta_\mu(x)+\varphi_\mu(x))\cos\phi_\mu(x),\\\nonumber
U_\mu^1(x)&=&\sin\phi_\mu(x)\cos\chi_\mu(x),\\\nonumber
U_\mu^2(x)&=&\sin\phi_\mu(x)\sin\chi_\mu(x),\\\nonumber
U_\mu^3(x)&=&\sin(\theta_\mu(x)+\varphi_\mu(x))\cos\phi_\mu(x),
\end{eqnarray}
read
\begin{eqnarray}\label{tbc2}
U^0_\mu(x)=\Biggl\{
\begin{array}{l}
U^0_\mu(x)\cos\varphi-U^3_\mu(x)\sin\varphi~~~~$for$~x=\{x_0,x_1,L_s,x_3\}~$and$~\mu=2,\\
U^0_\mu(x)\hskip 3.8cm~$for other links$,
\end{array}\\\label{tbc3}
U^3_\mu(x)=\Biggl\{
\begin{array}{l}
U^0_\mu(x)\sin\varphi+U^3_\mu(x)\cos\varphi~~~~$for$~x=\{x_0,x_1,L_s,x_3\}~$and$~\mu=2,\\
U^3_\mu(x)\hskip 3.8cm~$for other links$.
\end{array}
\end{eqnarray}
The relations (\ref{tbc2}) and (\ref{tbc3}) have been implemented
into the kernel of the MC procedure in order to produce the
configurations with the magnetic flux $\varphi$. In this case the
flux $\varphi$ is accounted for in obtaining a Boltzmann ensemble
at each MC iteration.

\section{Description of simulations and data fits}
The MC simulations are carried out by means of the heat bath
method. The lattices $2\times 8^3$, $2\times 16^3$ and $4\times
8^3$ at $\beta=3.0$, $5.0$ are considered. These values of the
coupling constant correspond to the deconfinement phase and
perturbative regime. To thermalize the system, 200-500 iterations
are fulfilled. At each working iteration, the plaquette value
(\ref{plaq}) is averaged over the whole lattice leading to the
Wilson action (\ref{Wilson}). Then the effective action is
calculated by averaging over the 1000-5000 working iterations. By
setting a set of magnetic fluxes $\varphi$ in the MC simulations
we obtain the corresponding set of values of the effective action.
The value of the condensed magnetic flux $\varphi_{min}$ is
obtained as the result of the minimization of the free energy
density (\ref{enactions}).

The spontaneous generation of magnetic field is the effect of
order $\sim g^4$ \cite{SB}. The results of MC simulations show the
comparably large dispersion. So, the large amount of the MC data
is collected and the standard $\chi^2$-method for the analysis of
data is applied to determine the effect. We consider the results
of the MC simulations as observed ``experimental data''.

The effective action depends smoothly on the flux $\varphi$ in the
region $\varphi\sim 0$. Therefore, the free energy density can be
fitted by the quadratic function of the flux $\varphi$,
\begin{eqnarray}\label{parfit}
\label{ffit} F(\varphi)=F_{min}+b(\varphi-\varphi_{min})^2.
\end{eqnarray}

This choice is motivated also by the results obtained already in
continuum field theory \cite{DS} where it was determined that free
energy has a global minimum at $\varphi\not=0$. The
parametrization (\ref{parfit}) is the most reasonable in this
case. It is based on the effective action accounting for the
one-loop plus daisy diagrams \cite{DS},
\begin{eqnarray}\label{diagr}
F(H)=\frac{H^2}{2}+\frac{11}{48}\frac{g^2}{\pi^2}H^2\log
\frac{T^2}{\mu^2}-\frac13\frac{(gH)^{3/2}T}{\pi}
-\frac{1}{12}\Tr\left[\Pi_{00}(0)\right]^{3/2},
\end{eqnarray}
having $g^2$ and $(g^2)^{3/2}$ orders in coupling constant. Here,
$H$ is field strength (flux $\varphi\sim H$), $T$ is the
temperature, $\mu$ is the normalization point, $\Pi_{00}(0)$ is
the zero--zero component of the gluon polarazation operator
calculated in the external field at the finite temperature and
taken at zero momentum. The value of $\beta=3$, which was used,
corresponds to a deep perturbation regime. So, a comparison with
perturbation results is reasonable. The systematic errors in
fitting function (\ref{parfit}) could come from not taking into
account the high-order diagrams in (\ref{diagr}). However, as it
is well known \cite{Linde}, the lack of an expansion parameter at
finite temperature starts from the three-loop diagram
contributions that is of $g^6$ order and could not remove an
effect derived in $g^2$ and $g^3$ orders. As the finite-size
effects are concerned, in the present investigation we just made
calculations for two lattices $2\times 8^3$ and $2\times 16^3$ and
have derived the same results for the $\varphi_{min}$ (as it will
be seen below). A more detailed investigation of this issue
requires much more computer resources, which were limited.

There are 3 unknown parameters, $F_{min}$, $b$ and $\varphi_{min}$
in Eq.(\ref{ffit}). The parameter $\varphi_{min}$ denotes the
minimum position of the free energy, whereas the $F_{min}$ and $b$
are the free energy density at the minimum and the curvature of
the free energy function, correspondingly.

The value $\varphi_{min}$ is obtained as the result of the
minimization of the $\chi^2$-function
\begin{eqnarray}
\chi^2(F_{min},b,\varphi_{min})=\sum_i\frac{(F_{min}+b(\varphi_i-\varphi_{min})^2-F(\varphi_i))^2}{D(F(\varphi_i))},
\end{eqnarray}
where $\varphi_i$ is the array of the set fluxes and
$D(F(\varphi_i))$ is the data dispersion. It can be obtained by
collecting the data into the bins (as a function of flux),
\begin{eqnarray}
D(F(\varphi_i))=\sum_{i\in
bin}\frac{(F(\varphi_i)-\hat{F}_{bin})^2}{n_{bin}-1},
\end{eqnarray}
where $n_{bin}$ is the number of points in the considered bin,
$\hat{F}_{bin}$ is the mean value of free energy density in the
considered bin. As it is determined in the data analysis, the
dispersion is independent of the magnetic flux values $\varphi$.
The deviation of $\varphi_{min}$ from zero indicates the presence
of spontaneously generated field.

\begin{figure}[t]
\begin{center}
\includegraphics[bb=11 70 776
538,width=0.4\textwidth]{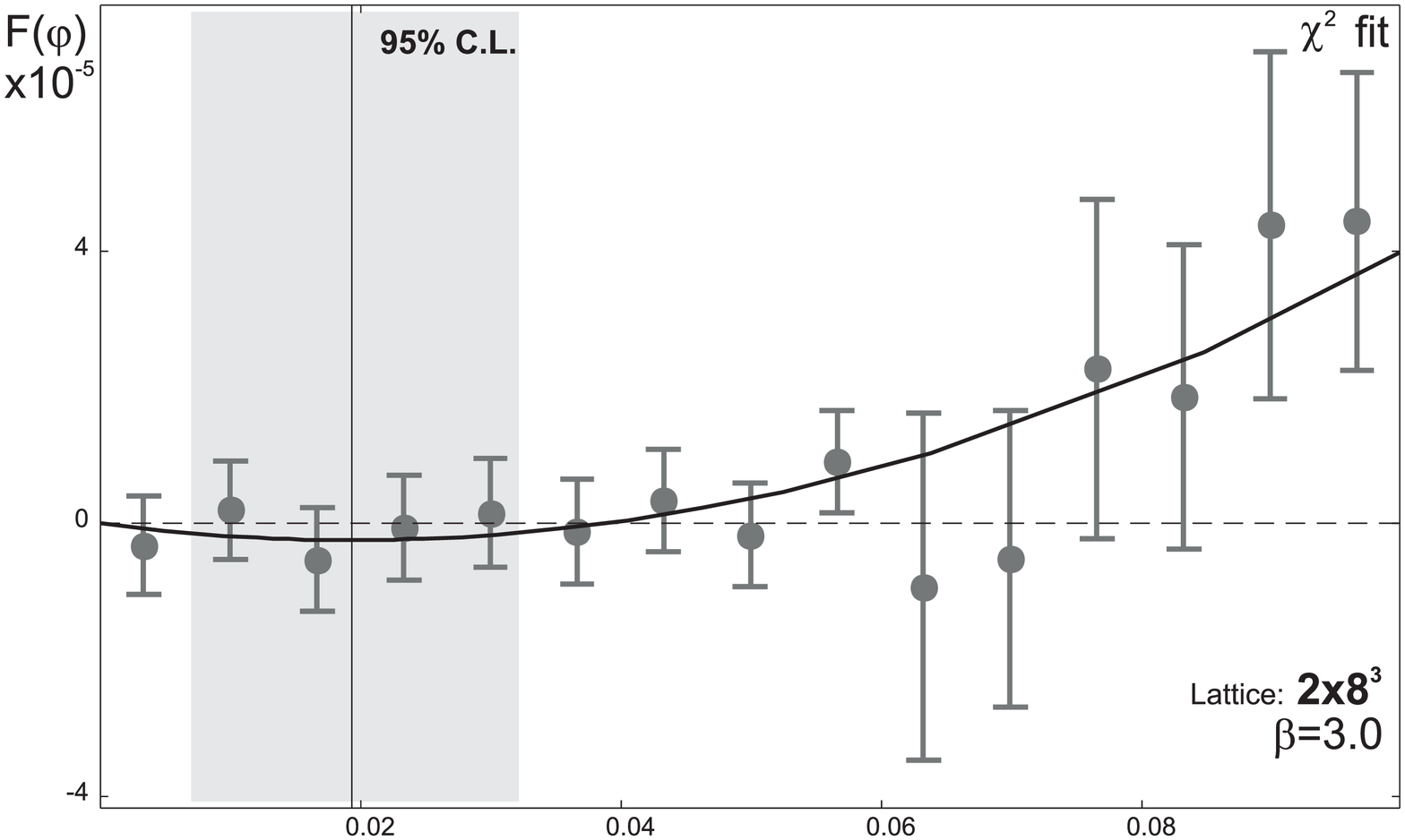}\hskip 1cm
\includegraphics[bb=20 108 802 567,width=0.4\textwidth]{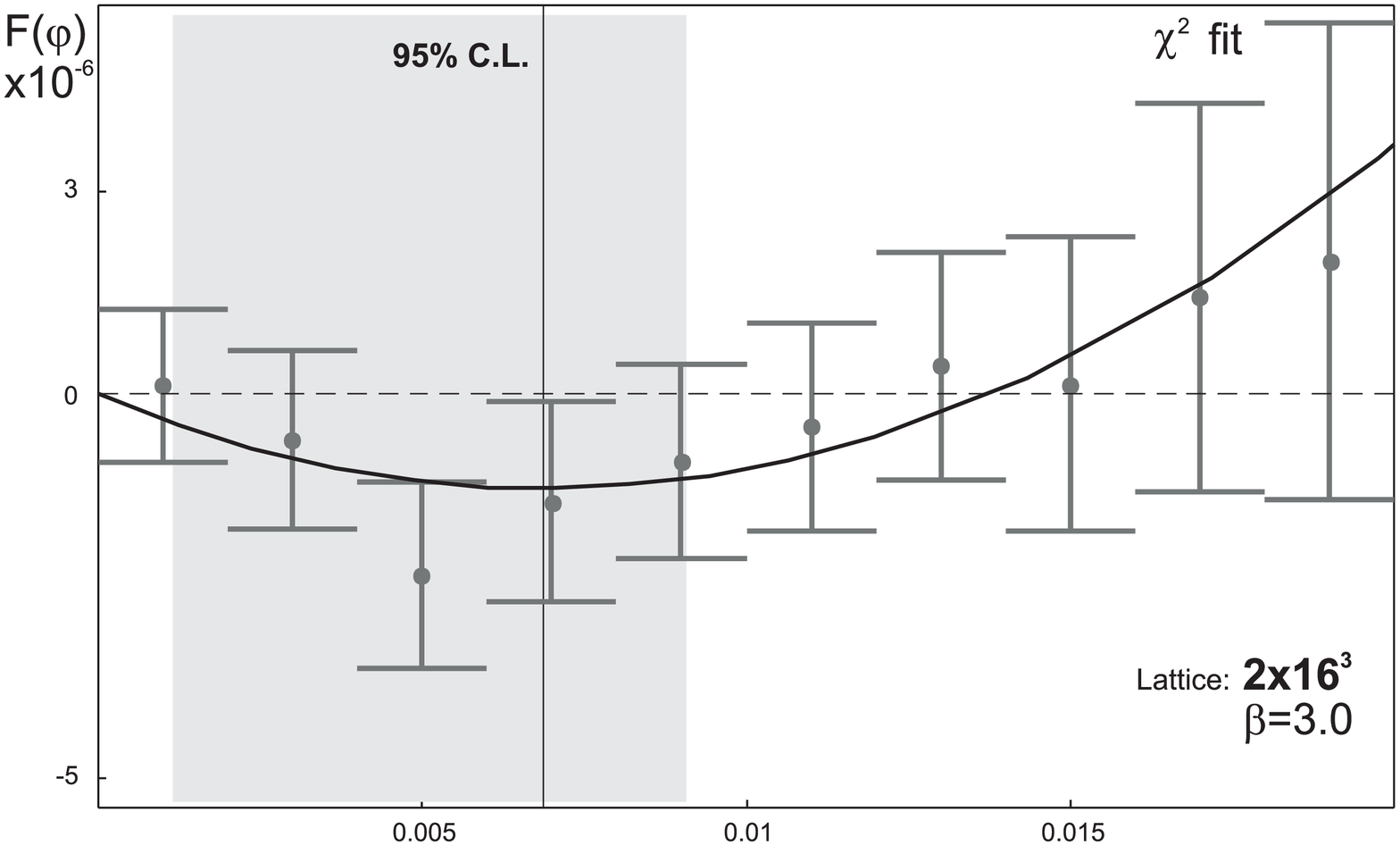}
\end{center}
\caption{$\chi^2$-fit of the free energy density on lattice
$2\times 8^3$ (left figure) and $2\times 16^3$ (right figure) for
{\small $\beta=3.0$} (grey regions describe the {\small
$\varphi_{min}=0.019^{+0.013}_{-0.012}$ and
$\varphi_{min}=0.0069^{+0.0022}_{-0.0057}$}, at the 95\% C.L.,
correspondingly).}
\end{figure}
\begin{table}[b]
\caption{The values of the generated fluxes {\small
$\varphi_{min}$} for different lattices (at the $95\%$ C.L.).}
\begin{center}
\begin{tabular}{|c|c|c|c|}
  \hline \rule{0pt}{14pt}
  & $2\times 8^3$ & $2\times 16^3$ & $4\times 8^3$ \\\hline
 \rule{0pt}{14pt}
  $\beta=3.0$ & ${0.019^{+0.013}_{-0.012}}$ & $0.0069^{+0.0022}_{-0.0057}$ & $0.005^{+0.005}_{-0.003}$ \\\hline
 \rule{0pt}{14pt}
  $\beta=5.0$ & ${0.020^{+0.011}_{-0.010}}$ &  &  \\\hline
\end{tabular}
\end{center}
\end{table}

The fit results are given in the Table 1. As one can see,
$\varphi_{min}$ demonstrates the $2\sigma$-deviation from zero.
The dependence of $\varphi_{min}$ on the temperature is also in
accordance with the results known in perturbation theory: the
increase in temperature results in the increase of the field
strength \cite{SB}.

The fits for the lattices $2\times 8^3$ and $2\times 16^3$ at
$\beta=3.0$ are shown in Fig. 2. The maximum-likelihood estimate
of $F(\varphi)$ by the whole data set is shown as the solid curve.
In addition, all $\varphi$ values are divided into 15 bins. The
mean values and the 95\% confidence intervals are presented as
points for each bin. The first 9 bins contain about 600-2000
points per bin. The large number of points in the bins allow to
find the free energy $F$ with the accuracy which substantively
exceeds the dispersion, $\sqrt{D(F(\varphi_i))}\sim 10^{-4}$. It
makes possible to detect the effect of interest. As it is also
seen, the maximum-likelihood estimate of $F(\varphi)$ is in a good
accordance with the bins pointed, because the solid line is
located in the 95\% confidence intervals of all bins.

The 95\% C.L. area of the parameters $F_{min}$ ($b$ for the right
figure) and $\varphi_{min}$ is represented in Fig. 3. The black
cross marks the position of the maximum-likelihood values of
$F_{min}$ ($b$ for the right figure) and $\varphi_{min}$. It can
be seen that the flux is positive determined. The 95\% C.L. area
becomes more symmetric with the center at the $F_{min}$, $b$ and
$\varphi_{min}$ when the statistics is increasing. This also
confirms the results of the fitting.
\begin{figure}[t]
\begin{center}
\includegraphics[bb=123 157 520 543,width=0.3\textwidth]{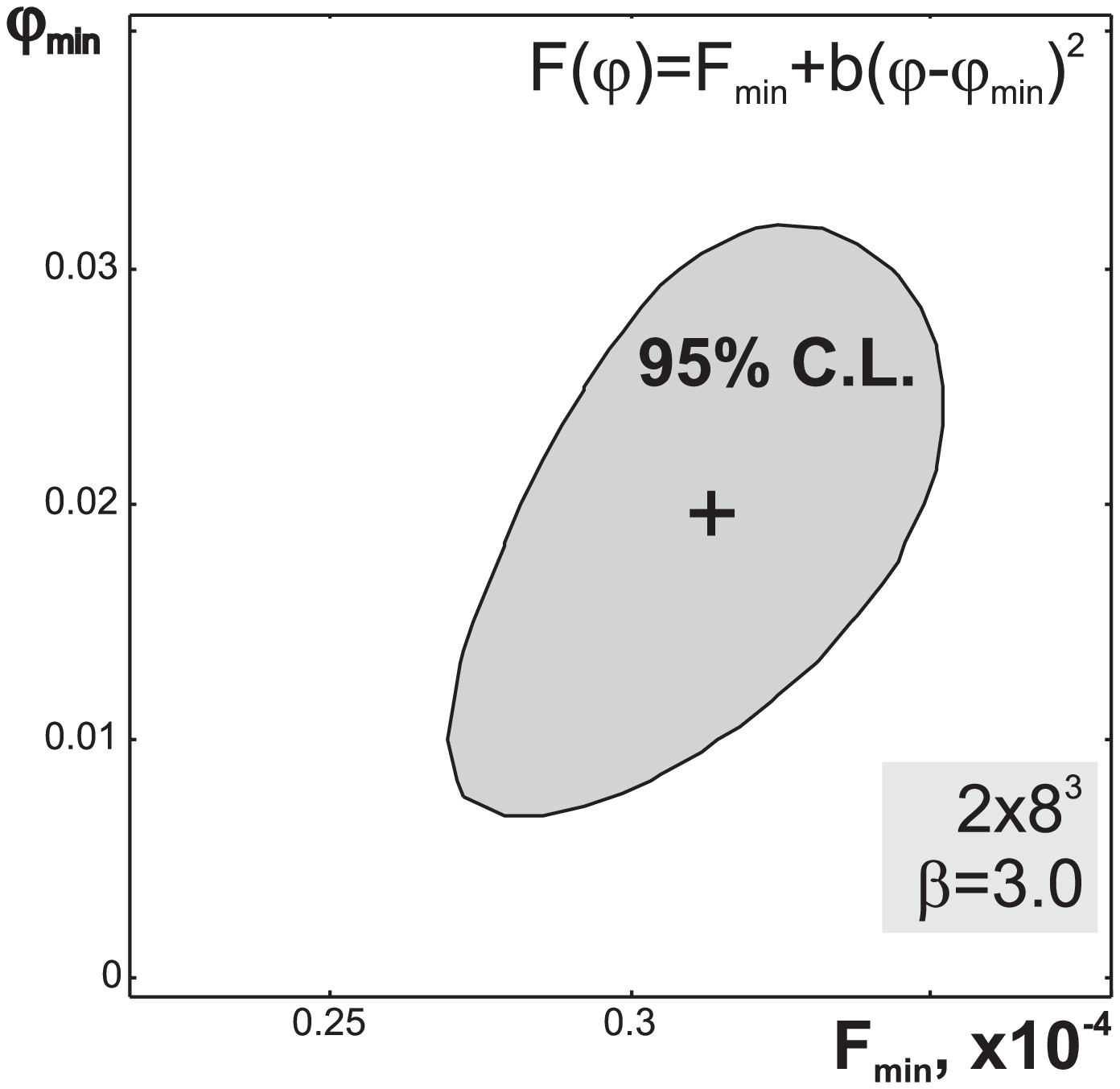}\hskip 2cm
\includegraphics[bb=86 294 471 665,width=0.3\textwidth]{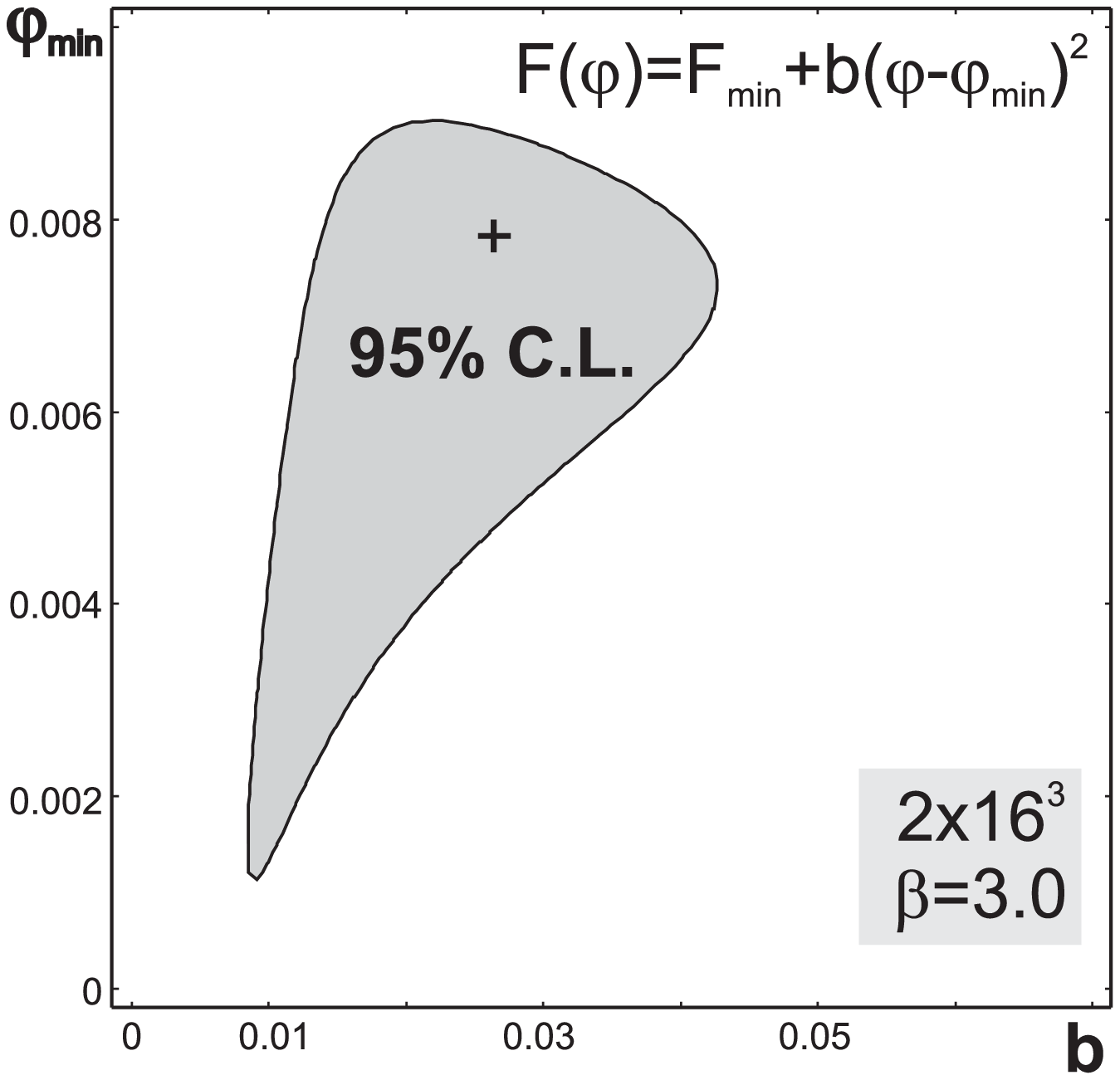}
\end{center}
\caption{The 95\% C.L. area for the parameters $F_{min}$ and
$\varphi_{min}$, determining the free energy density dependence on
the flux $\varphi_{min}$ on lattice $2\times 8^3$ for $\beta=3.0$
(left figure). The 95\% C.L. area for the parameters $F_{min}$ and
$b$ , determining the free energy density dependence on the flux
$\varphi_{min}$ on lattice $2\times 16^3$ for $\beta=3.0$ (right
figure).}
\end{figure}

\section{Discussion}
The main conclusion from the results obtained is that the
spontaneously created temperature-depen\-dent chromomagnetic field
is present in the deconfinement phase of QCD. This supports the
results derived already in the continuum quantum field theory
\cite{SB,SS} and in lattice data analysis \cite{AG}.

Let us first discuss the stability of the magnetic field at high
temperature. It was observed in Refs. \cite{SB,SS} that the
stabilization happens due to the gluon magnetic mass calculated
from the one-loop polarization operator in the field at
temperature. This mass has the order $m^2_{magn} \sim g^2
(gH)^{1/2} T \sim g^4 T^2$ as it should be because the
chromomagnetic field is of order $(gH)^{1/2} \sim g^2 T$
\cite{SB}. The stabilization is a nontrivial fact that, in
principle, could be changed when the higher order Feynman diagrams
to be accounted for. Now we see that the stabilization of the
field really takes place.

Our approach based on the joining of calculation of the free
energy functional and the consequent statistical analysis of its
minimum positions at various temperatures and flux values. This
overcomes the difficulties peculiar to the description of the
field on a lattice. Here we mean that the field strength on a
lattice is quantized and therefore a nontrivial tuning of the
coupling constant, temperature and field strength values has to be
done in order to determine the spontaneously created magnetic
field.

We also would like to note that in the present paper the flux
dependence on temperature remains not investigated in details.
This is because of the small lattice size considered. That
restricts the number of points permissible to study. However, at
this stage we have determined the effect of interest as a whole.
Even at the small lattice, one needs to take into consideration
thousands points of free energy (that corresponds to an analysis
of 5-10 millions MC configurations for different lattices) to
determine the flux value $\varphi_{min}$ at the $95\%$ C.L. In
case of larger lattices this number and corresponding computer
resources should be increased considerably. This problem is left
for the future.

As we mentioned in Section 2, the finite-size effects have not
been investigated in detail. However, these effects are important
near the phase transition temperature. They make difficult to
distinguish a first-order phase transition from a second-order
one. In our case, the temperature is far from $T_c$. The fact that
external field penetrates the Coulomb phase is well known
\cite{CC,Forcr}, so the only really new thing is that this field
is spontaneously created. It was first observed in continuum
\cite{SB}, where the field strength of order $g^4$ in coupling
constant was established. Finite-size effects are not able to
remove this result. The values $\varphi_{min}$ obtained on the
lattices $2\times 8^3$ and $2\times 16^3$ (see the Table 1) are in
a good agrement with each other, within the statistical errors at
$95\%$ C.L.

One could speculate that the lattice sizes $2\times 8^3$ and
$2\times 16^3$ are not sufficient. However, these lattice sizes
were used in the Refs. \cite{DeGrand}. The main aim of present
paper is to show a possibility of spontaneous generation of
chromomagnetic field at high temperature in lattice simulation,
which was investigated already by perturbative methods
\cite{SB,S2}.

It is interesting to compare our results with that of in Ref.
\cite{CC} where the response of the vacuum on the external field
was investigated. These authors have observed in lattice
simulations for the $SU(2)$- and $SU(3)$-gluodynamics that the
external field is completely screened by the vacuum at low
temperatures, as it should be in the confinement phase. With the
temperature increase, the field penetrates into the vacuum and,
moreover, increase in temperature results in existing more strong
external fields in the vacuum. On the other hand, increase in the
applied external field strength leads to the decreasing of the
deconfinement temperature. These interesting properties are
closely related to the studies in the present work. Actually, we
have also investigated the vacuum properties as an external field
problem when the field is described in terms of fluxes. This was
the first step of the calculations. The next step was the
statistical analysis of the minimum position of free energy, in
order to determine the spontaneous creation of the field. In fact,
at the first step we reproduced the results of Refs. \cite{CC} (in
terms of fluxes).

Note that the present investigations also correspond to the case
of the early universe. They support our previous results on the
magnetic field generation in the standard model \cite{DS} and in
the minimal supersymmetric standard model \cite{DS2}. As it was
discussed by Pollock \cite{PL}, the field generated by this
mechanism at the Planck era might serve as a seed field to produce
the present day magnetic fields in galaxies.

We would like to conclude with the note that the deconfinement
phase of gauge theories is a very interesting object to study. The
temperature dependent magnetic fields, which are present in this
state, influence various processes that should be taken into
consideration to have an adequate concept about them.

\vskip 1cm

\section*{Acknowledgement} The authors would like to express
sincere gratitude to Michael Ilgenfritz for his kind attention and
help at each stage of the work. We also thank Alexey Gulov for
numerous useful discussions. One of us (VD) is indebted for
hospitality to ICTP (Trieste), where the final part of the work
has been done.

\end{document}